# Toward Sustainable Rare Earth Element Production: Key Challenges in Techno-Economic, Life Cycle, and Social Impact Assessment

Adam Smerigan[1] and Rui Shi[1,2]*

[1]Department of Chemical Engineering, The Pennsylvania State University, University Park, Pennsylvania 16802, United States

[2]Institute of Energy and the Environment, the Pennsylvania State University, University Park, Pennsylvania 16802, United States

*Corresponding author: rms6987@psu.edu

## Abstract

Rare earth elements (REEs) are 17 critical minerals used in many clean energy technologies like wind turbines and electric vehicles. Conventionally, we produce REEs from mining in few, geopolitically restricted regions. Developing systems that utilize new technologies and unconventional feedstocks provides an opportunity to meet increasing demand while improving sustainability. Techno-economic analysis (TEA), life cycle assessment (LCA), and social LCA (sLCA) are commonly used tools to assess the sustainability performance of these systems. However, analyses of REE systems encounter challenges including system scope, data availability, technology scale-up, and uncertainty. In the reviewed literature, systems served multiple functions beyond producing REEs, including circularizing production and waste remediation, leading to discrepancies in scope. Further, the instability of REE prices led to high uncertainty due to different revenue, costs, functional unit, and impact allocation. Therefore, these analyses leave decision makers with an incomplete understanding of the current landscape of REE production inhibiting intelligent and efficient identification of future direction. In this narrative review, we conducted a comprehensive overview of the literature, synthesized studies from each pillar of sustainability (economic, environmental, and social), highlighted the challenges and limitations in each field, and recommended direction for future work developing sustainable REE production systems.

# 1 Introduction

To enable clean energy transitions and meet our climate goals,[1] we must secure sustainable sources of critical minerals including REEs. The demand for REEs (consisting of Sc, Y, and the lanthanides) is expected to increase by 3-7 times by 2040.[2] This increase in production is unlikely to be filled by traditional mining efforts, which are both technically inefficient and geopolitically restricted. As a result, the future REE supply chain must integrate systems that utilize secondary sources using emerging technologies.[3] However, the sustainability performance of these novel systems must be rigorously validated. Without such an assessment, there is a risk of investing in solutions that are unprofitable, environmentally harmful, or socially inequitable.

Currently, REEs are primarily produced through five production routes at specific production sites: Bayan Obo (BO, China), southern provinces (SP, China), Sichuan province (SC, China), Mount Weld (MW, Australia), and Mountain Pass (MP, United States). For Chinese production from bastnaesite/monazite ores (BO, SC) and ion adsorption clays (SP), the entire process (from mining through refining) is completed domestically. For production routes in Australia and the United States, REEs from bastnasite ore are mined and extracted into a concentrated REE mixture before being shipped internationally (to Malaysia and China, respectively) for separation and refining. In total, China is estimated to produce 85% of unrefined REEs and 95% of all refined REEs.[4–6] These current production routes have several negative aspects: geopolitical tensions affecting market stability, inefficient intra-REE separations,[7,8] and impacts on local communities due to mining operations.[9] Further, conventional mining may be unable to meet increased demand since opening additional mines can be a long and unsuccessful process.[10] As a result, other production routes are being explored that use secondary REE feedstocks to reduce dependence on geographically concentrated virgin material mining, eliminate industrial wastes, and circularize the rare earth supply chain.

It is critical to assess the sustainability of these novel systems at early stages of development. To this end, several methods have been used to assess the environmental, economic, and social aspects of sustainability (LCA,[11,12] TEA,[13–15] and sLCA,[16,17] respectively). However, characterizing the tradeoffs across all three pillars of sustainability (e.g., profit vs. global warming vs. equity) is challenging. Further, there are methodological limitations in assessing REE systems that make conclusions about system sustainability uncertain. These issues are exacerbated by low quality data and insufficient uncertainty and sensitivity analyses. Hence, it is difficult to identify system hotspots for further investigation and prioritize targets for developing technologies.

Without standardized and comprehensive sustainability analyses, we risk inefficient technology development and diminished incentives for innovative REE production systems—both of which are crucial for supplying modern technologies and securing climate security into the future. To advance a clearer understanding of the sustainability of emerging REE production pathways, this review article aims to accomplish the following objectives: (1) synthesize knowledge within each discipline of sustainability assessment, (2) assess challenges and limitations within these works, (3) provide resources and recommend improvements.

We structure the main body of this text into sections to discuss the current landscape for REE production, the limitations of analyses for each pillar of sustainability, and our recommendations based on this review. These results will help researchers standardize sustainability assessment methodologies in the field and enable consistent comparisons between novel REO production systems and conventional systems. Overall, this work will support the development of sustainable REE production systems that are economically viable, environmentally responsible, and designed to meet global needs while contributing to the circular economy.

## 2 Current Process Landscape for REE Recovery

Several flowsheets have been explored for secondary feedstocks including post-consumer products (e.g., catalysts, magnets), post-industrial waste (e.g., coal fly ash, acid mine drainage), and other minerals (e.g., lignite coal, phosphate rock).[18–20] However, these feedstocks each have unique compositions of REEs and impurities that make REE extraction and purification challenging. Therefore, new technologies are being developed (e.g., bioleaching[21,22], electrochemical processes[23,24], sorptive separations[25,26]) to reduce the cost and environmental burden of these systems. Nevertheless, low technological readiness makes it difficult to assess the performance of these technologies at full-scale for comparison with conventional routes.[27,28] Trying to use technological data from conventional REO production from REE containing ores is challenging since these data are proprietary.[29] Nevertheless, certain information, including simple process flowsheets and cost data, is available in the literature.[29,30] These simple flowsheets and fundamentals have been leveraged for alternative feedstocks. These routes generally follow Figure 1 where a solid feedstock is beneficiated followed by leaching of REEs into the liquid phase before removing impurities and selectively separating individual REEs for further refinement. Figure 1 also highlights some of the primary established and developing technologies used in REE production from a variety of feedstocks. Other reviews have been published for specific feedstocks and technologies.[8,21,31]

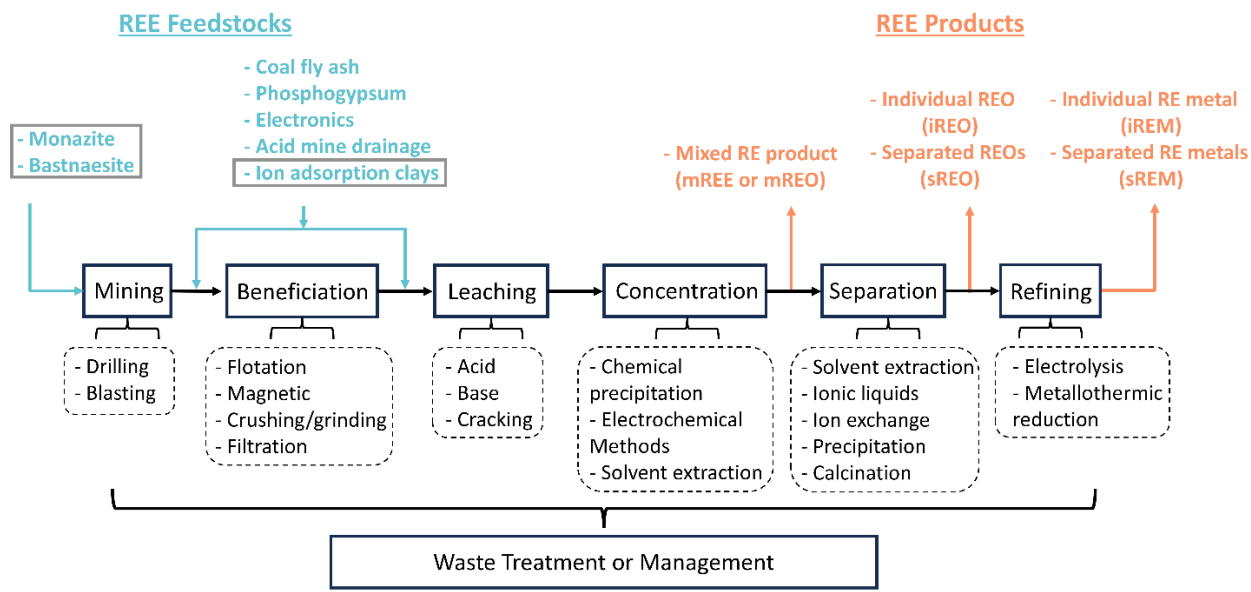

Figure 1: System diagram of rare earth element (REE) production from feedstock acquisition through refining. Different feedstocks are shown in blue text with conventional sources being boxed in gray. Process sections are shown in black boxes with potential technologies in dotted boxes. Different possible rare earth products are shown in orange text.

### 2.1 Economics and Environmental Impact of Conventional REE Production

A previous study[30] has compiled extensive economic data for conventional REE production processes. This review reported a large variance in the cost of operation for different REE production facilities. The capital expenditure (CAPEX) ranged from $0.016 to $608 per kilogram of REE produced (in 2022 US dollars) and the operational expenditure (OPEX) ranged from $0.91 to $11,000 per kilogram of REE produced. On average, REE production costs approximately $27.83·kg$^{-1}$ REE. These costs are further broken down by process section with the REE-selective separation and refining comprising the majority of process costs (67% of the cost from mining through refining).[30] The revenue of these REE production facilities is a function of the composition of the REEs in the feedstock as well as the REE purity. The price per kilogram of the sum of REEs produced is referred to as the basket price, which has varied widely over the past 20 years. A typical basket price has been reported to range from approximately $15 to $55·kg$^{-1}$ REE, though higher basket prices were observed around 2011 due to policy changes in China.[32]

For the environmental impact of REE production, there have been many studies using a variety of different assumptions and uncertainties. Here, we summarize the reported global warming (GW) impact to quantify the collective uncertainty of several conventional REO production studies. The results for BO and SP have been multiplied by an adjustment factor that accounts for illegal mining that occurs in these locations (S1). Figure 2a shows that the GW impact varies nearly an order of magnitude across different studies for BO and SP routes. This high uncertainty makes it less clear whether novel systems have lower GW impact than conventional systems. Figure 2b shows how the GW contribution of process sections varies between the different conventional production routes.

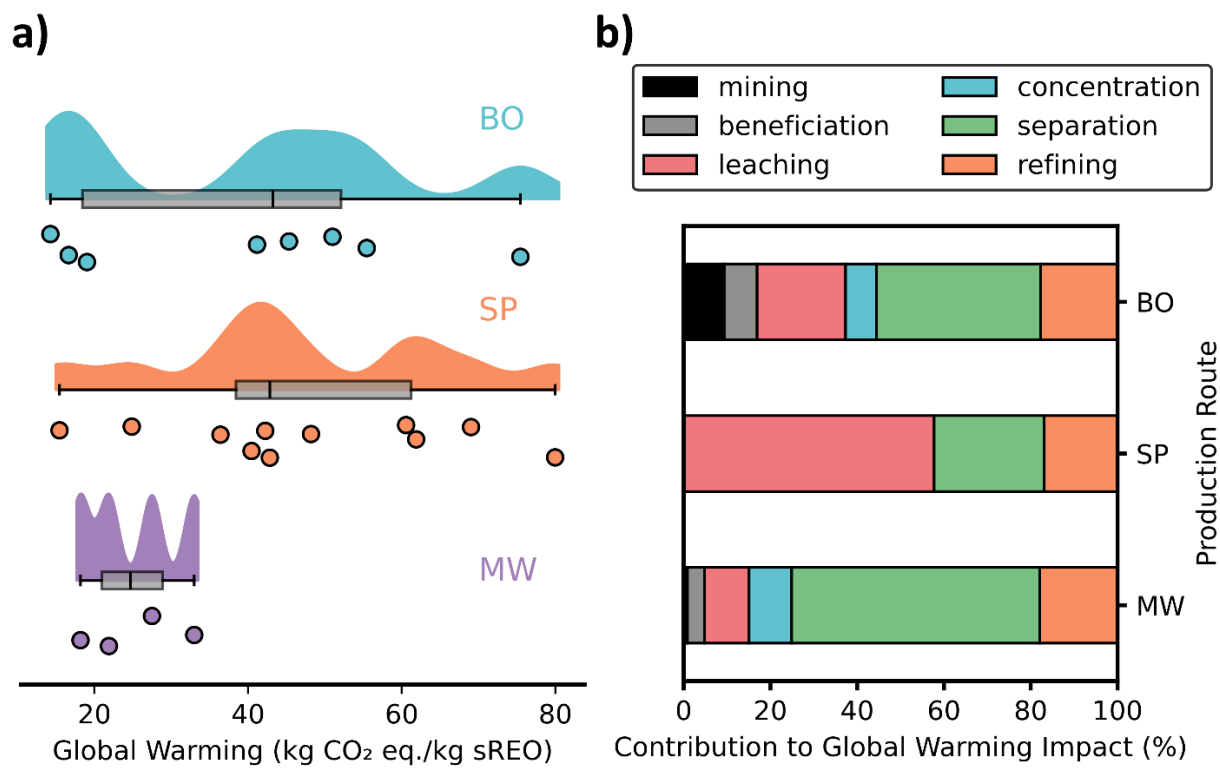

Figure 2: Compiled literature data for the global warming impact of conventional REE production routes (Bayan Obo China (BO), Southern Provinces China (SP), and Mount Weld Australia (MW)) per kilogram of separated REOs. This data is broken down into **(a)** the total global warming impact of each REE production route where each dot is a data point, the shaded regions are kernel density plots, and the gray is a boxplot and **(b)** the contribution of each process section to global warming impact.

# 3 Limitations and Gaps in Present Sustainability Assessment Practices

## 3.1 Techno-economic Analysis (TEA)

In the reviewed literature, TEA studies investigated post-industrial waste (e.g., coal waste), the recycling of post-consumer wastes (e.g., NdFeB magnets), and ores. The REE content across feedstocks varied considerably from less than 0.001 wt% in acid mine drainage up to 50 wt% in monazite ore. Further, the most abundant REE varied by feedstock: cerium was the most prevalent in ores and post-industrial wastes (around 20-35 wt% REE), Nd in NdFeB magnets (63-89 wt% REE) and cell phones (70 wt% REE), and La in FCC and NiMH batteries (97-100 wt%). In 15 of the 16 studies, the REE production capacity was at least 1 mt·year$^{-1}$ with more abundant feedstocks having larger capacities. However, one study on NdFeB magnets[33] considered a capacity less than 1 mt·year$^{-1}$ leading to very unfavorable economics as compared to other NdFeB studies.[34–36] Within these 16 studies, overall REE recovery varied (from 20-100 wt%) depending on feedstock composition and selected processing technologies. Depending on the feedstock, different contaminants and concentrations of REEs influenced leaching efficiency[33] and solvent extraction performance.[30] Overall, there is great diversity in the composition and availability of each feedstock leading to technological challenges and distinct design decisions. Additionally, there were inconsistencies in cost estimations regarding whether waste treatment and storage units are included in TEA.

We summarize these TEA studies in Table 1 and discuss their results and limitations in five sections: 1) product identification, 2) product valuation, 3) operating costs (OPEX), 4) capital costs (CAPEX), and 5) profitability analysis. Here we highlight some of the main points and variability in TEAs of REE systems, with methodological discussions following established TEA methodologies and industry practices.[14,28]

### 3.1.1 Complexity in Product and Coproduct Identification.

Table 1 shows the products across different feedstocks and system designs. Potential REE products consist of a mixed concentrate (mREE as oxide, chloride, mischmetal, etc.), individual REEs (iREE as oxide, carbonate, etc.), reduced RE metal (REM), or as a highly processed product (REEp, magnets). Generally, products that are more pure have significantly improved value (with some variability due to the type of REE).[30] However, further purification incurs additional costs. Therefore, potential incorporation of additional downstream processing must balance increasing revenue with cost. In 11 of the 16 studies, most of the revenue comes from only one REE, indicating that purification of the other REEs may not be worth the expense. Some studies have considered whether a mixed product has sufficient purity and composition to be directly used without incurring the expense of further separation (e.g., a Nd/Pr/Dy mixture for NdFeB magnets[34,37,38] and a La/Ce mixture as mischmetal[20]). Other studies have found that coproducts are the main source of revenue. These coproducts range from avoided waste tipping fees (FCC waste at 97% of revenue[22]), other metal products (NiO at 79%[39] and Au at 85%[40] of revenue), and other chemicals (phosphoric acid at 71% of revenue[41]). The revenue from these coproducts also helps insulate system profitability from the high variability of REE market values. Nevertheless, the identification and valuation of coproducts has been inconsistent across studies, and insufficient detail is reported regarding the value added from coproducts to determine optimal strategies.

### 3.1.2 Limitations and Inconsistencies in REO/REE Product Valuation.

Beyond identifying the REE products, it is challenging to estimate the value of each REE product. Many studies use the "basket price", which evaluates the sREO value using the value of each individual REE within the feedstock.[32] Basket prices were commonly reported in the literature for feedstocks containing multiple REEs (e.g., monazite, coal fly ash, AMDp). For studies that didn't explicitly list a basket price, we calculated the basket price to directly compare product values between feedstocks (Table 1). Basket prices varied widely from as little as $2/kg to over $400/kg in some cases. However, most basket prices varied between $50-150/kg. The high price ($1038/kg) for the geothermal brine study is due to producing solely a pure, high-value Eu product.

However, there are limitations to using the basket price. Basket prices are based on market data which is limited and highly variable (up to 400% fluctuation in REO values since 2010[42]). Market prices are available primarily as >99.9% pure individual oxides (iREOs) or metals (REMs) making it difficult to assess the value of unpurified, mixed REE (mREE) products. To overcome this limitation, studies have used several methods to determine the value of REE products with varying accuracy. The least accurate method assumed iREO and mREE products have the same value. Another method applied discount factors to iREO values to estimate mREE values (assume the value of mREEs is around 20-60% the value of iREOs[30,43]). Another method added a toll cost for using an external solvent extraction processes to convert their mREE product into an iREO product and iREO values were used to calculate revenue.[44] Lastly, product value correlations were developed from available data to predict the value of REE products as their purity increases (from mREE to REEp).[30] However, the REE balance problem may lead to deflation in the prices of more abundant REEs that are coproduced with desired, high-demand REEs.[3] Therefore, operations with low operational expenses are expected to perform better than those with higher basket prices for REOs.[32] In summary, there is high uncertainty in the value of REE products of varying purity making it difficult to quantify revenue and compare across different systems.

Table 1: Summary of the TEA studies reviewed for each feedstock with relevant feedstock information.

| | Post-consumer products | | | | | | Post-industrial waste | | | | Mined ores | |
|---|---|---|---|---|---|---|---|---|---|---|---|---|
| | Fluid cracking catalyst | NiMH batteries | Hard disk drive | Cell phone | NdFeB magents | Lamp phosphors | Geothermal brine | Coal fly ash | Acid mine drainage | Phosphoric acid sludge | Monazite | Lignite coal |
| Reviewed studies (#) | 1 | 1 | 1 | 1 | 4 | 1 | 1 | 2 | 3 | 1 | 1 | 1 |
| References | 22 | 39 | 37 | 40 | 33–36 | 36 | 45 | 46,47 | 38,44,48 | 41 | 49 | 46 |
| Plant Capacity (mt/year) | 18,800 | 2,000 | 342 | 2000 | 0.33-2,270 | 2610-11,000 | 1340 | 3,650-200,000 | 749-4,400,000 | 453,000 | 16,000 | 200,000 |
| REE production (mt/year) | 137 | 245 | 2.53 | 32 | 0.095-689 | 421-2410 | 2.15 | 1.1-47.1 | 1-444 | 49-138 | 7110-7290 | 83 |
| REE recovery (wt %) | 28-56 | 85 | 75 | 78 | 60-100 | 66 | 90 | 45-75 | 80-93 | 17-48 | 80-90 | 75 |
| REE content (wt %) | 1.5 | 12 | 0.7-1 | 1.6 | 22-36 | 18-31 | 0.16 | 0.003-0.09 | 0.11-0.9 | 0.09 | 50 | 0.06 |
| Primary REE (wt %) | La 97 | La 100 | PrNd 100 | Nd 70 | Nd 63-89 | Y 90-92 | Eu 100 | Ce 33-36 | Ce 25-26 Y 30 | Ce 27 | Ce 25 | Ce 32 |
| Basket Price ($/kg REO) | 2-101 | 14 | 150 | 35 | 38-110 | 31-51 | 1038 | 22-104 | 24-104 | 256 | 14.5 | 491 |
| Primary REE (% of revenue) | La 92-97 | La 100 | PrNd 100 | Nd 92 | Nd 36-82 Dy 51 Tb 84 | Tb 97-99 | Eu 100 | Nd 37 Sc 70-90 | Dy 18-23 Sc 80 | Lu 73 | Nd 50 | Sc 70-94 |

### 3.1.3 Uncertainty and Variability in Operational Expense (OPEX)

From the literature reviewed, we broke down the OPEX into five distinct categories: materials, utilities, labor, maintenance, and other fixed costs (Figure 3). The cost of maintenance and other fixed costs were typically estimated as a percentage of other capital or operating costs. Labor costs were primarily determined by multiplying the number of shift workers (determined based on number of process sections and phase of materials) by salary. To calculate materials and utilities costs, several studies took prices from literature and online sources (e.g., Alibaba and mineralprices.com) and multiplied these prices by the rates of consumption for each item. However, prices listed on those websites can vary widely, lack verification, and may not reflect industrial-scale procurement costs. Using more standardized databases, such as the ICIS database,[50] or supplier quotes would strengthen the quality of OPEX calculations and enable more consistent comparisons between studies. Prices from previous years were often converted to present value using inflation rates or price indices. Another key limitation observed across several studies is the lack of consideration of waste treatment or management. Of the 16 studies reviewed, only 7 mentioned these expenses in the text. Among these 7 studies, the reported costs either varied significantly or were not specified at all—some studies merely acknowledged their inclusion without providing exact figures.

Overall, material costs were the largest contributor to OPEX (mean of 51%) and had the largest variance (4-88%). This large variance is due to different feedstocks and technologies having different chemical requirements (e.g., whether acid leaching is required). The next largest contributor was other fixed costs (mean of 21%), which also had a large variance (0-72%). This variance is due to different methodologies for calculating fixed costs. Utilities, labor costs, and maintenance costs were least influential on the OPEX (9%, 15%, and 4%, respectively) and had less variation with only one or two outliers on the high end. For the studies with high labor costs, one paper[35] was a batch process and the other paper[48] gave no details on how the number was estimated. For the study with a high utility cost,[33] high rates of material recycling reduced material costs leading to a higher contribution of utility costs to the total OPEX. Overall, material costs dominate in hydrometallurgical mining systems emphasizing the need for higher recycling rates and more efficient technologies that require fewer materials to extract and concentrate REE metals, whether through novel leaching technologies or more selective separations.

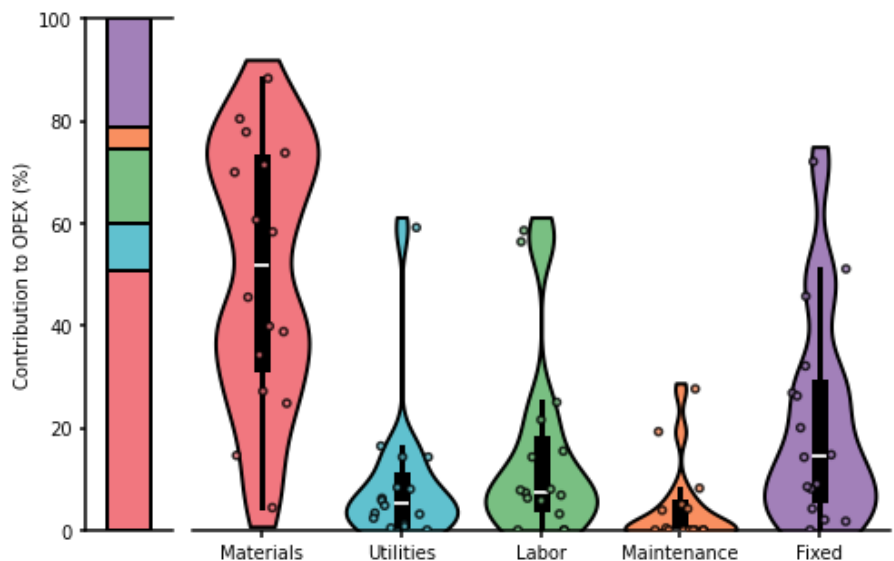

Figure 3: Contribution of different costs to total OPEX based on the average value from 16 literature sources (left) and the variation of the values from these different sources (right) where each dot is a data point, the colored area is a kernel density distribution surrounding a box plot where the median is a white line.

### 3.1.4 Inconsistent and Inadequate Capital Expenses Estimation Methods for REE Systems

Capital expenditure (CAPEX) is the sum of inside battery limit costs (e.g., f.o.b, equipment purchase costs (EPC) and installation costs), outside battery limit (OSBL) costs, engineering and construction costs, working capital, and contingency charges. For REE production systems in the literature, CAPEX estimation are performed using two methods: order-of-magnitude and study estimate methods (i.e., the Lang factor method). Using these methods, a TEA is expected to have ± 50% and ± 35% accuracy for the order-of-magnitude and study estimate methods, respectively.[14,51] However, uncertainty can be as little as 5-10% as TRL increases and detailed purchase and installation cost estimates (e.g., vendor quotations) are obtained.

Beyond this inherent uncertainty, there is variability in the way CAPEX was estimated in the studies reviewed here. These studies used several methods to collect the EPC including literature sources, websites, and size factor calculations in process design textbooks. If purchase costs were obtained from smaller equipment, costs were scaled to size using the 6/10 rule.[14] Other studies used size factors and process engineering textbook cost correlations to estimate full-scale equipment costs. It was unclear whether the EPC costs accounted for delivery costs. After calculating purchase costs, most studies converted prices that were obtained in past years to present day values using either an assumed inflation rate or cost indices (e.g., the Chemical Engineering Plant Cost Index, CEPCI). However, 8 of the 16 studies made no distinction as to the cost basis of their study. After calculating the present value of scaled up equipment, it was unclear how most studies calculated the installed cost of the equipment. Many studies mentioned using the Lang factor method but did not provide sufficient detail to understand what factor they used. Other studies made no mention of how installed costs were calculated, if they were calculated at all. Beyond the inside battery limits costs, the extent to which "other investment costs" were estimated was highly variable. Of the 16 studies, 9 provided "other costs" (e.g., contingency or working capital) that were used in the final estimate of CAPEX. However, each study had its own list of "other costs" that ranged up to an order of magnitude between studies.

To better understand how CAPEX was estimated in these studies and to facilitate the standardization of methods in this emerging field, we calculated the ratio of total CAPEX to EPC. In Figure 4, these studies clustered into two distinct groups around ratios of 2 and 6. Most of the systems with high CAPEX:EPC ratios had similar REO production capacities and CAPEX to the other studies. These high ratio studies cited Towler and Sinnott's[15] method for their CAPEX estimations suggesting that this method may lead to higher CAPEX:EPC ratios. Higher CAPEX:EPC ratios agree more closely with generally accepted lang factors which predict CAPEX:EPC ratios around 4-6 depending on the phases of flows in the system.[14] However, these lang factors are based on decades-old data (the most up-to-date Lang factors available are from 2003[13]) and may no longer be accurate, affecting the accuracy of CAPEX estimates. New developments in CAPEX estimation methods are needed to better reflect the characteristics of such emerging and novel processing.

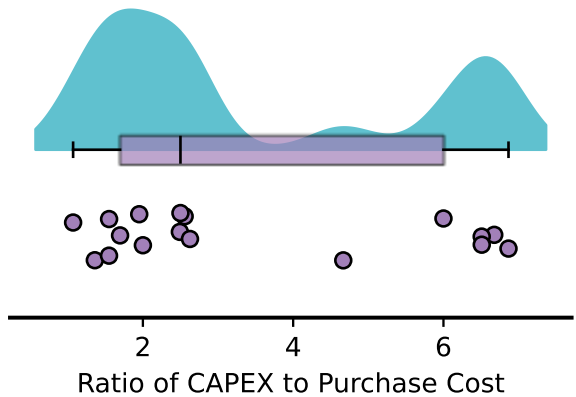

Figure 4: The ratio of capital expenditure to equipment purchase cost for each of the studies reviewed (purple dots). The statistics and distribution of the data are represented by a box plot (purple) and kernel density plot (blue).

### 3.1.5 Transparency Issues in Profitability Evaluation and Lack of Discounted Cash Flow Analysis

To assess the economic viability of systems, economic indicators are compared between existing and proposed systems. Some indicators do not account for the time value of money (e.g., return on investment, payback period). Other indicators, such as net present value (NPV) and internal rate of return (IRR), consider the time value of money through a discounted cash flow (DCF) analysis. Of the 16 studies reviewed, 11 completed some form of DCF analysis, while the others reported only revenue and costs. For the calculation of NPV, studies used discount rates between 3-10% (8 of these 11 studies used rates between 8-10%). IRR was infrequently reported in the literature. Beyond discount rate, there are several other important parameters in DCF analysis. Studies considered plant lifetimes from 10-40 years with the average being between 20-30 years. Nine of the 16 studies included construction periods and 5 studies included loan repayment schedules for CAPEX. Most studies included depreciation over 7-20 years using a variety of depreciation methods (e.g., straight line). For more details, the values of these parameters are summarized in S3.

Of the 16 papers reviewed, 5 used an incomplete approach to profitability assessment by not including a DCF analysis, limiting both the accuracy of their evaluation and comparability across studies. For the studies reviewed, we directly report the NPV for studies that completed a DCF analysis. To compare across all studies, we calculated two metrics that do not include a DCF analysis: the return on investment (ROI) and cash flow (a common metric used in this field). The cash flow was calculated as the difference between annualized revenue and the annualized CAPEX and OPEX (equations provided in S5). Further, to consider a range of scenarios, we compiled the reported best- and worst-case values for given ranges of revenue and OPEX. For CAPEX, we used the average value for all scenarios to simplify the analysis since the range and magnitude of CAPEX values were small. Therefore, a study that gave ranges of revenue and OPEX would have 4 permutations in these results. These results in

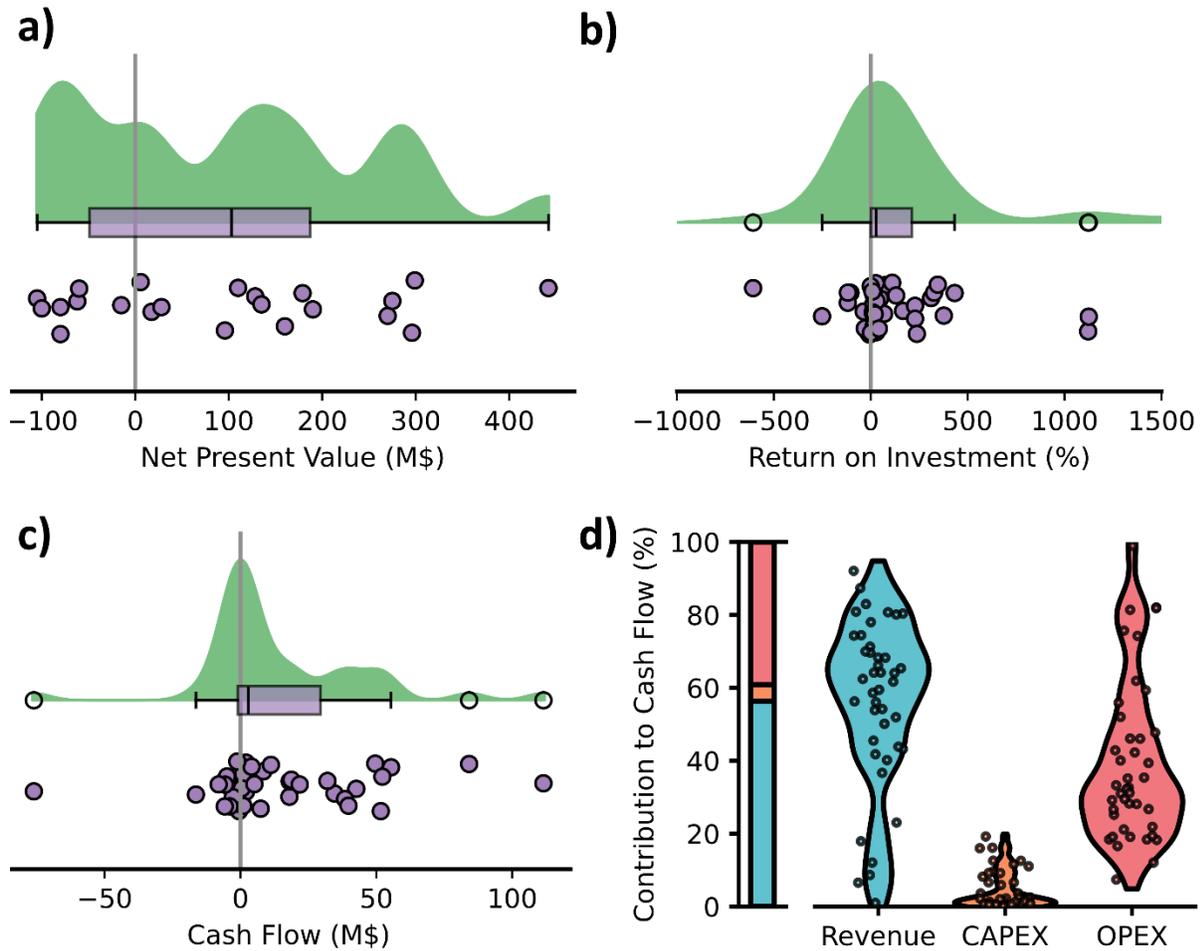

Figure 5a-c show that approximately 72% of systems were profitable across a range of different feedstocks and scenarios, regardless of whether the DCF analysis was completed. However, the return on investment (ROI) and cash flow metrics do not account for the time value of money. Therefore, these metrics are less useful for comparing between different investments over dissimilar project lifetimes.

To understand the main driver of profitability in these studies, we broke down the cash flows into revenue, annualized CAPEX, and OPEX for these systems (

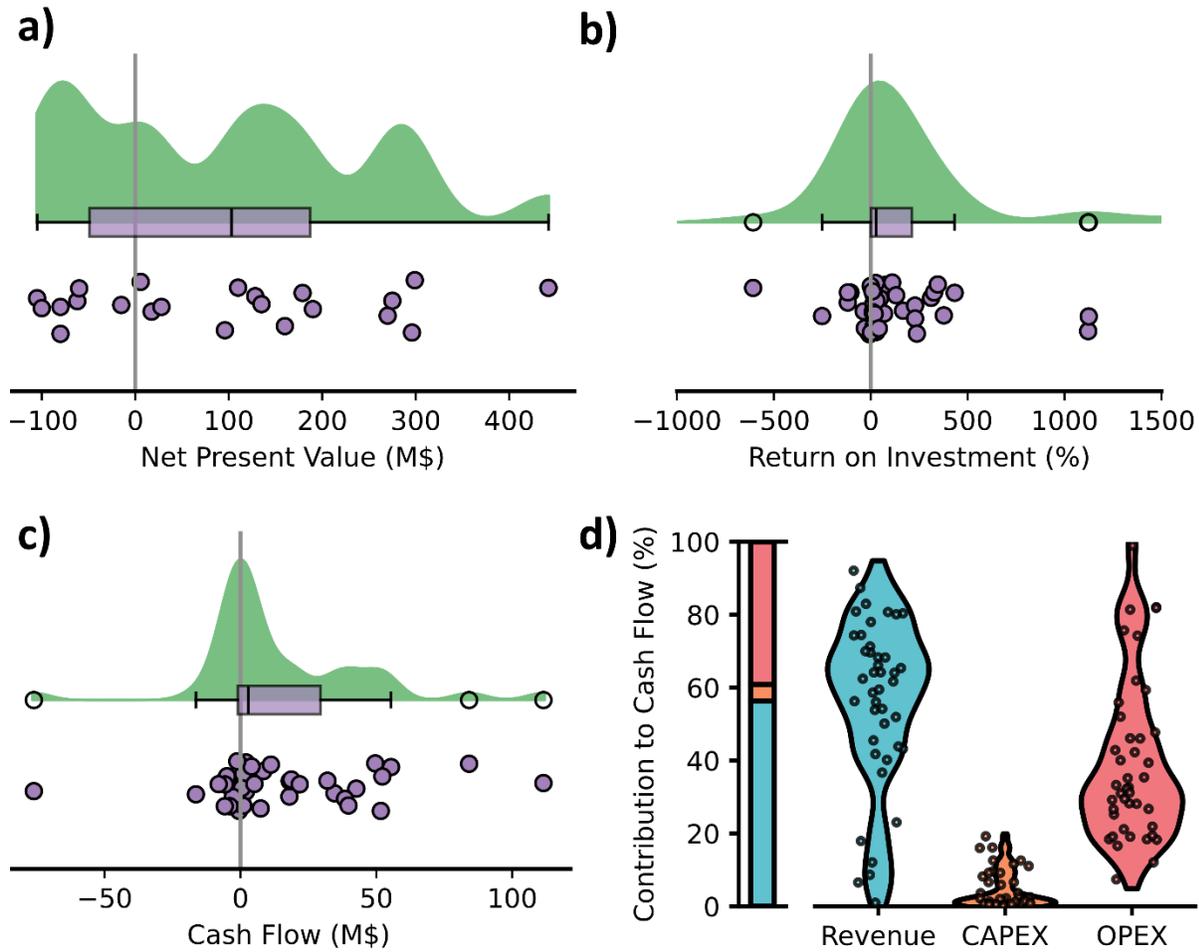

Figure 5d). Again, we compiled this data across different revenue and OPEX scenarios to understand how different factors (e.g., prices, feedstock composition, technology efficiency, etc.) may affect profitability. Overall, the CAPEX had limited impact on the profitability of these systems. The OPEX contributed almost as much to the cash flow as revenue (36% and 59%, respectively). This result suggests that systems with low OPEX will be more resilient to fluctuating REE values and are most likely to succeed, which agrees with other works.[30,32,44] Unfortunately, achieving lower OPEX can be challenging for unconventional feedstocks with low REE content, as there is additional expense in extracting and concentrating these REEs. Therefore, technological innovations will be key for accessing low REE content feedstocks to compete with higher grade feedstocks.

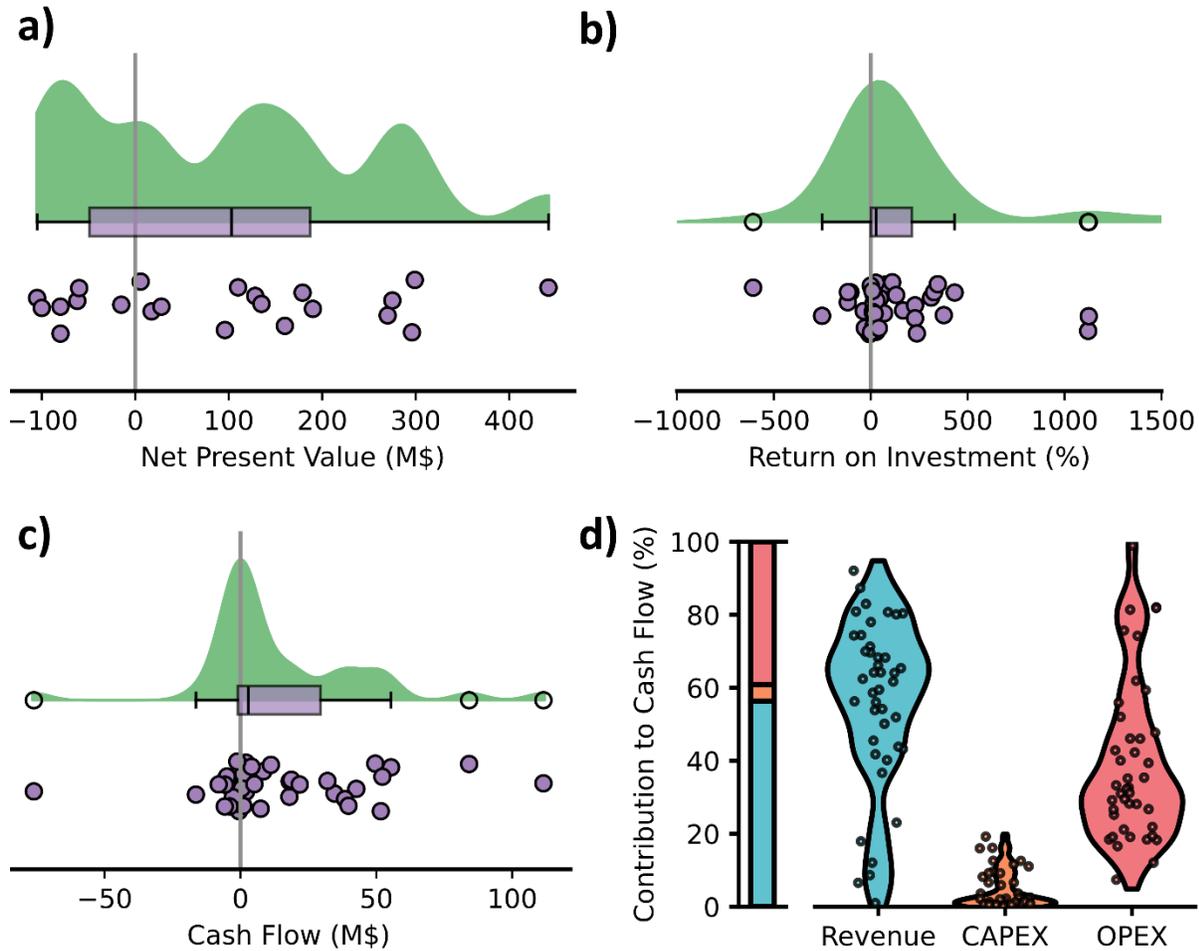

Figure 5: The profitability of REE production systems as **(a)** net present value, **(b)** return on investment, and **(c)** cash flow. **(d)** The total cash flow broken down into the contribution of annualized revenue, annualized CAPEX, and OPEX. The bar represents the mean of each contributor while the dots and violin plots show each sample and its distribution from the reviewed studies and scenarios.

## 3.2 Life Cycle Assessment (LCA)

To assess environmental impact, the most common method is life cycle assessment (LCA). LCA guidelines exist through international standards (ISO 14040 and 14044)[11,12] making LCA more reputable, widely used, and comparable between studies. However, there are several issues limiting the reliability and usefulness of LCA for REO production technologies. Herein, we summarize the life cycle impacts of conventional REO production for assessing the life cycle environmental impact of novel REO production systems. We summarize the studies reviewed in Table 2.

Additionally, we break down the limitations of LCAs for systems producing REEs into several groups: 1) functional unit and scope, 2) multifunctionality and allocation, 3) life cycle inventory uncertainty, and 4) life cycle impact assessment uncertainty. Ideally, this work will bring to light some deficiencies of LCA studies for REO production and lead to more robust LCAs with clearer (or at least more informed) conclusions for decision makers.

Table 2: Summary of LCA studies reviewed including the reference, date of publication, feedstock, functional unit, LCIA method, allocation method, and co-products produced. If mREEs are separated into sREO products, an "X" is marked in the "separate" column. If sREOs are refined into sREM products, an "X" is marked in the "refine" column.

| Feedstock[1] | Functional Unit | LCIA Method | Allocation[2] | Coproduct | Separate | Refine | Date | References |
|---|---|---|---|---|---|---|---|---|
| BO, SP | iREO, iREM, REEp | TRACI v2.1, ILCD | E | Iron ore, REOs | X | X | 2018 | 52 |
| BO, SP, MW | sREO | ILCD v1.09 | E, S | Iron ore | X | | 2020 | 53 |
| BO, MW, MP | REEp | ReCiPe v1.08, H | E, M | Iron ore, REOs | X | X | 2018 | 54 |
| BO | mREO | TRACI | E, M, O | Iron ore, REOs | | | 2015 | 55 |
| SP | mREO | Impact 2002+, USEtox 2.01, IPCC | - | - | | | 2019 | 56 |
| BO, SP | iREO, sREO | CML 2002 | E, M | Iron ore, REOs | X | X | 2017 | 57 |
| BO, SP, SC | sREM | Eco-costs, others | - | Iron ore | X | X | 2018 | 58 |
| BO, SP, NK | iREO, sREO | ReCiPe v1.08, H | E, S | Iron ore, REOs | X | X | 2018 | 59 |
| all | iREO | n.a. | n.a. | Iron ore, REOs | X | X | 2022 | 60 |
| SP | mREO | TRACI, ILCD | E | - | | | 2016 | 61 |
| BO, SP | mREE | TRACI, ILCD | E | REOs | X | | 2017 | 62 |
| SP | iREF, iREM | TRACI | E | REOs | X | X | 2018 | 63 |
| BO, NK | iREM | ReCiPe v1.08, H | E, S | Iron ore, REOs | X | X | 2016 | 64 |
| BO, SP | iREO, sREO | - | - | - | X | | 2021 | 65 |
| BO | iREO | EI99 | E, M | Iron ore, REOs | X | | 2014 | 66 |
| MW | iREO, sREO | EI99, EI95 | E | REOs | X | | 2020 | 67 |
| SP | sREO | CML baseline v4.4 | - | - | X | | 2017 | 68 |
| BO, hard disk drives | REEp, sREO | CML 2001 | E | Iron ore, REOs | X | X | 2014 | 69 |
| Idaho soil | Soil processed | CML-IA baseline | - | - | | | 2024 | 9 |
| Fluorescent lamp powder (FLP) | FLP processed | EF 3.0 | D | mREO | | | 2021 | 70 |
| Phosphogypsum | PG treated | EI 99, H/A | D | anhydrite, H3PO4 | | | 2016 | 71 |
| Coal refuse, lignite | mREO | TRACI v2.0 | E | REOs | | | 2020 | 46 |
| Coal refuse, AMD | mREOH, sREO | TRACI | - | - | X | | 2024 | 72 |
| NdFeB magents | mREO | TRACI v2.1, CED | E | Iron oxide | | | 2021 | 34 |
| NdFeB magents | REEp | TRACI | - | - | | | 2016 | 73 |
| e-waste | Gold | TRACI, ILCD | E | Silver, copper, mREO | | | 2019 | 74 |
| Hard disk drives | mREO | TRACI | E, M | Iron salt | | | 2024 | 37 |
| NdFeB magents | REEp | EF v3.0 | - | - | | X | 2024 | 75 |
| LED waste | mREO, lighting service | ReCiPe H. endpoint | M, D | mercury | | | 2020 | 76 |
| Fluid cracking catalyst | FCC waste processed | TRACI v2.1 | - | Disposal credit | | | 2018 | 22 |
| NdFeB magents | mREO | CML-IA baseline v3.05 | E, M | - | | | 2022 | 33 |

[1]Feedstocks for conventional mining are listed by their production route: monazite/bastnaesite at Bayan Obo (BO), ion adsorption clays in Southern Provinces, China (SP), bastnaesite at Mountain Pass, United States (MP), and eudialyte at Norra Kärr, Sweden (NK).
[2]Allocation methods used including economic (E), mass (M), other (O), subdivision (S), displacement (D).

### 3.2.1 Functional Unit and System Boundary

Functional units are the foundational component of environmental assessments, providing a basis for quantifying the performance of different systems or products (an apples-to-apples comparison). However, within the 31 studies reviewed, there is a distinct lack of clarity for the specific product(s) being produced, which leads to confusion in interpreting the functional unit. One study goes as far as describing all REE compounds as REOs by stating the following: *"…for the sake of clarity, we often refer to all of these different forms of rare earths as rare earth oxides (REO)"*.[69] As we continue to advance sustainable solutions for REE recovery, we must carefully select functional units to enable meaningful comparisons across different approaches. In Figure 1, we show a generic system diagram for REO production from primary and secondary sources, illustrating how different REE feedstocks enter the system at various process sections due to the varying extractability of REEs. Depending on the chosen scope of a study, five different functional units can be chosen (orange text). The first product exits the system after the concentration section as a mostly pure mixed REE product (mREE). This product can be a combination of multiple REOs (e.g., mischmetal, or other non-product composition) or REEs associated with another ion (e.g., RE carbonate or chloride). With further processing (e.g., solvent extraction), the mREEs can be separated into individual REEs. At this point, two different functional units can be defined: the mass of an individual purified REO (iREO, like $Nd_2O_3$) and the mass of separated REOs (sREO, the sum of iREOs after separation). This choice has implications for allocation discussed in Section 3.2.2. Further refinement of the sREOs through molten salt electrolysis or metallothermic reduction produces a pure RE metal (REM). Again, two functional units can be chosen for either individual (iREM) or separated RE metals (sREM, the sum of iREM produced). Table 2 highlights the stark contrast in scope and functional unit between conventional routes and unconventional feedstocks. Specifically, looking at the separation column, only one study of an unconventional feedstock considers the selective separation in the scope of the study illustrating that apples-to-apples comparisons to conventional systems is challenging.

To bridge this gap, we developed a method to enable comparisons between different functional units. In Figure 6, we use the GW impact of a mREE producing system to estimate the impact of separating and refining mREEs into sREO and sREM products, thereby converting between these functional units. The conversion from mREE to sREM requires four numbers: (1) the GW impact for a 100% pure mREE product (divide by purity if mREE has impurities), (2) the percentage of the original amount of REEs from the mREE in the final REE product (Figure 6d), (3) the fraction of impacts from the separation step (Figure 6e or Figure 2), and (4) a conversion to account for the mass of oxygen removed during refining. Figure 6b and c illustrate how to recalculate the GW impact based on the targeted functional unit (further details and limitations included in S4). By developing approaches to compare systems with different functional units, we can enable rapid comparison of environmental impacts between systems to enhance decision-making.

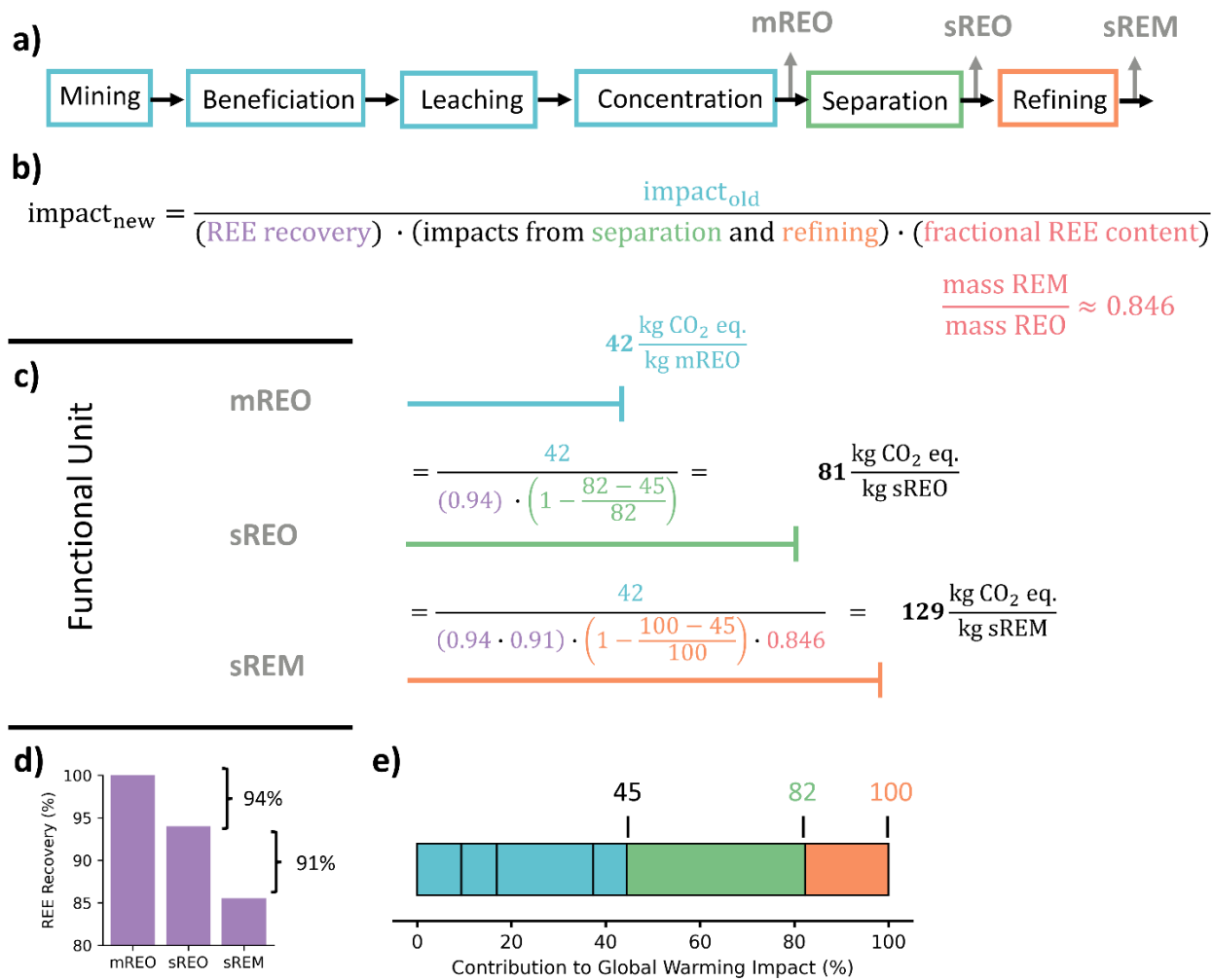

Figure 6: One example of how to convert between functional units used in REE production studies. This example considers that a study (with a global warming impact of 42 kg $CO_2$ eq/kg mREO) wants to estimate the total global warming impact after separation and refining of that mREO product into sREO and sREM. **(a)** shows a flowsheet to produce the different REE products broken down by process sections. **(b)** the generic equation for converting between functional units with each term color coded for clarity. **(c)** shows the three functional units that can be converted between (mREO, sREO, and sREM). **(d)** shows the average amount of REEs recovered after additional processing where 100% of REEs are initially present as the mREO product. **(e)** shows, on average, how much each process section contributes to the total impact of producing REMs for the Bayan Obo processing route.

### 3.2.2 Multifunctionality and Allocation

ISO standards[11,12] describe how systems with multiple functions or products distribute, or allocate, environmental impacts among various outputs of a system. In the context of REE recovery, both conventional and unconventional feedstocks produce coproducts in addition to REEs (e.g., iron ore at Bayan Obo). Therefore, it is necessary to decide how to allocate process burdens to these co-products. For iron and niobium ore coproduced at BO, a two-step economic allocation has been performed to attribute some of the impacts to these products.[69,77] The influence of this decision has not been rigorously explored. In other cases, when the coproduct is highly valuable (e.g., electronics with trace REO, copper, and gold), the impacts from REEs can range orders of magnitude (or even become negative) depending on the allocation procedure used. Therefore, it is important to carefully consider how to allocate impacts to both coproducts and amongst individual REEs. Below, we step through the ISO standards and summarize the different allocation approaches used in the LCAs of REE systems and their limitations.

For certain functional units, such as 1 kg of mREE produced or 1 kg sREO produced, allocation between individual REEs is not required. This approach is simple and convenient, making comparisons between different systems easy at first glance. However, different feedstocks and processing technologies lead to differing compositions of REE products. For example, ion adsorption clays have a higher composition of heavy REEs than monazite and bastnaesite. Heavy REEs are typically less abundant and have higher value compared to light REEs. This discrepancy can complicate comparisons, as all forms of "1 kg of separated REOs" are not equivalent in their value and end use.

The ISO 14044 standard[12] first recommends avoiding allocation through system expansion and system subdivision. During separation, all REEs are dependent on the initial separation step that produces the first individual REE, but only the remaining REEs are dependent on further downstream separations. Therefore, subdivision has been used in some studies to reflect this partial independence.[54,64] However, subdivision is questionable due to the physical (e.g., shared solvent) and economic dependence of the separation on each REE.[53,78] In the literature reviewed, it was commonly unclear how the subdivision had been performed making it difficult to compare between studies. System expansion may also be inappropriate as the "displaced" processes may not exist in reality.[77]

Continuing along the ISO 14044 guidelines[12], impacts can be attributed to individual products by allocating based on physical relationships (e.g., mass, energy), or, if necessary, economic value. Within the REO literature reviewed, numerous allocation methods have been used to attribute impacts to iREOs.[54,55,79] However, the most commonly used allocation method is economic allocation, where the mass and price of each coproduct is used to determine how to distribute the impact. Several studies have shown how mass allocation can shift impacts to the more abundant light REEs while economic allocation shifts impact to heavy REEs, which are typically of higher value.[55,66] Therefore, if a functional unit of 1 kg of La oxide produced (a light REE) is used and economic allocation is performed, the system would have very low impact since most of the impact has been attributed to the heavy REE products. Many studies in the field do well to emphasize these limitations. However, a more involved discussion reporting multiple functional units and allocation methods is frequently not examined. Further, economic allocation relies on highly variable and uncertain REO prices, as described in Section 3.1.2. This variability will likely extend into the future, as REO demand is projected to increase (e.g., to supply the clean energy transition[2]), new feedstocks with different compositions of REOs are utilized.[3] Therefore, it is important that studies using allocation are fully transparent with their allocation factors to allow for modifications as the values of REOs change over time.

Replication of studies using subdivision or other allocation methods is often difficult or impossible without detailed methodological descriptions and transparent data. Even for the widely utilized ecoinvent database, it was challenging to convert the reported impacts (per kg of iREO) to per kg of sREO. Using ecoinvent's methods, a mass-based subdivision is performed for reference products followed by economic allocation to both reference products and allocatable byproducts. Though this method is common practice within ecoinvent, the methodology is not easily accessible nor is the ecoQuery documentation[80] sufficiently descriptive to replicate these results. EcoQuery does make clear whether REOs are reference products or allocatable byproducts, or explain the basis for their classification, which does not appear to rely on mass or market value.[80] Since allocated impacts for REEs depends on these designations, the impact results may be ambiguous, lacking consistency or methodological rigor. Further, since reverting this allocation is challenging, it is difficult to quantify this uncertainty and to compare between different functional units. To improve transparency, we have included an Excel file of the ecoinvent allocation calculations (provided in the SI document 2). Overall, the problems identified in the ecoinvent dataset are just one example of how multifunctionality is reducing the reliability of LCA results and their utility in assessing novel REE production systems.

### 3.2.3 Limited Data Availability for Life Cycle Inventories (LCI)

Many of the challenges associated with LCI collection for conventional systems are covered in two recent reviews.[53,65] For one, primary data availability (from the actual processes) for conventional REO production routes is sparse with few studies having direct measurements from industry. Therefore, many studies have used assumptions and estimations to fill data gaps (e.g., using regulatory information for toxic waste management). Some mistakes have been found in these calculations (e.g., blasting) and in the implementation of assumptions (e.g., REE content of ores).[65] Many studies have used parts of previous works in their LCIs leading to few truly distinct LCIs. The use of similar foundational assumptions could artificially decrease the variation in results, leading to a false sense of accuracy. Over time, studies have begun to fill these gaps (e.g., solvent extraction), but primary data is still needed. Further, there is significant illegal production of REOs, where even less is known about these systems (as much as 40% of overall production from IAC in China).[58] Illegal mining impacts were estimated to be around 215% of the impact of conventional mining (S1). In addition to limited site-specific data, background LCI data has been missing for some flows. A few of these chemicals include $P_2O_4$ (used in solvent extraction)[77] and oxalic acid (used in REE precipitation)[34,55]. In more recent studies, the impact of these chemicals has proven significant.[60,62] Overall, despite over 15 LCAs for conventional REO production, there is large uncertainty in the LCAs of conventional systems that has not been quantified by rigorous uncertainty analysis.

For novel systems from secondary sources, many of these REE extraction and processing technologies are still in early stages of development, where full-scale primary data does not yet exist. Therefore, performing LCAs on emerging technologies has several challenges including comparability, scale, data, and uncertainty.[27] Issues with comparability relate to system boundary and functional unit, as discussed in Section 3.2.1. Additionally, these novel systems primarily rely on lab-scale data, which likely overestimates impacts compared to optimized, full-scale systems. The data required to model these steps accurately may be sparse or inconsistent due to the technical complexity of REE extraction and separation (e.g., bioleaching or sorption) from unconventional sources. As these new technologies develop, background databases may not have the chemicals needed for these new or specific technologies requiring the use of proxy chemicals. Therefore, these novel systems may have large uncertainty making it challenging to compare to conventional systems and to identify future direction for technology development.[27]

### 3.2.4 Limitations of the Current Impact Assessment Methodology

Life cycle impact assessment (LCIA) methods are used to quantify the impact of flows to the environment for specific impact categories. For a minority of impact categories (e.g., global warming, ozone depletion), the methods to quantify environmental impacts are largely consistent due to standardized frameworks established by the International Panel on Climate Change (IPCC)[81] and World Meteorological Organization (WMO)[82]. For the other impact categories, impacts from different methodologies can vary substantially, even if they employ the same units, due to different assumptions that influence characterization factors. Therefore, it is difficult to compare between studies that use different LCIA methods. In the REE literature reviewed, many LCIA methods have been used including CML, ILCD, TRACI, and ReCiPe. Further, within each LCIA method, there have been different versions which have their own impact categories and characterization factors.

When choosing an LCIA methodology to use, it is essential to consider all relevant impacts categories to understand all potential tradeoffs. For REO production, some impact categories are frequently reported (e.g., global warming, acidification, ecotoxicity). However, some other relevant, but less developed, impact categories are irregularly examined: radioactivity, water use, and long-term impacts of waste storage. Below, we discuss the limitations of current LCIA methodology in quantifying impacts in these categories.

### 3.2.4.1 Impacts from Radioactivity

During the processing of ores and industrial wastes containing REEs, commonly radionuclides (e.g., Uranium, U and Thorium, Th) entire the ecosphere from fine particle and fluid releases. However, quantifying the impacts of these radionuclides has two main challenges. One challenge is the lack of data on the amount and concentration of radionuclides crossing the system boundary. For conventional REE systems, radioactive wastes are produced during beneficiation, leaching, and extraction when the radionuclides are extracted from the feedstock alongside the REEs. The end fate of these metals is generally long-term, acid mine waste storage sites. These waste sites have varying levels of management where there is the possibility of toxic metals leaching into the environment. Some notable examples of the human impacts of this waste management are examined in Section 3.5. Beyond quantifying the radioactive flows, it is difficult for LCIA methods to determine the impact of these flows. Some LCIA methods (e.g., TRACI v2.1) [83] do not contain common flows from mining like U and Th. Other LCIA methods (e.g., ReCiPe 2016)[84] consider radioactive flows. However, the methodology used to quantify the impact of these radioactive flows is still developing. First, radionuclide releases are commonly irregular discharges (e.g., break in containment), which is typically not considered when conducting LCAs. Further, after these releases, it is difficult to model how radionuclides move and accumulate in the environment, which can be dependent on geography. Beyond challenges modelling the movement of radionuclides through the environment, directly relating radiation exposure to human health impact is difficult since long-term low-dose radiation can express as diseases now and in future generations.[85] Further discussion of the implications of radionuclide impacts for REE systems is discussed in other reviews.[53,65]

### 3.2.4.2 Water Use and Depletion Impacts

Since there are large amounts of water used in hydrometallurgical systems, water footprint should be considered in LCAs of REE production. However, there are inconsistencies in how water use impacts are modeled across LCIA methods.[75] One key detail is defining water use versus water depletion. Water use quantifies the amount of water used by a system, while water depletion is the amount of water consumed by the system that will not return to its original aquifer. When considering water depletion, a system that recovers REEs from wastewater while also treating that wastewater would receive a credit for returning water to the environment. However, in common LCIA methods, like ReCiPe 2016 (v1.1), it is unclear whether this treated wastewater is considered a credit within the system boundary. In the ReCiPe report[86], due to the complexity of linking water consumption to biosphere flows, implementation is left to interpretation. We examined implementations of ReCiPe water depletion from both openLCA and ecoinvent (v3.9.1). The openLCA implementation includes water flows back to the environment as a credit (characterization factor of -1), while ecoinvent's implementation does not. In wastewater treatment literature, it is acknowledged that most existing LCIA methods do not account for water flows back to the environment, which is important for water reclamation systems.[87] Their recommendation is to use the LCIA method called AWARE (Available WAter REmaining),[88] which links flows to water depletion impact. However, this LCIA method is not implemented in ecoinvent making it challenging to perform the analysis in some programs. Understanding the fate of water in REE production systems (e.g., indefinite storage as acidic tailings) is critical for managing the water, food, energy nexus, especially in geographies with water scarcity.

### 3.2.4.3   Impacts of Long-term Waste Storage

In the United States, 20% of the mining facilities inspected by the environmental protection agency (EPA) between 1990-1995 violated regulations from the Clean Air Act (CAA), Clean Water Act, and Resource Conservation and Recovery Act (RCRA) due to mismanagement of acid drainage and tailings disposal.[89] Further, secondary REE feedstocks (e.g., coal fly ash) and wastes from REE production are classified as Technologically Enhanced Naturally Occurring Radioactive Materials, which are regulated under the CAA, National Emission Standards for Hazardous Air Pollutants, and more recently the RCRA as a mixed waste.[90,91] Therefore, the environmental impact of these wastes must be quantified in LCAs of REE production systems from both primary and secondary feedstocks. However, there are currently no standardized methods, and it is unclear how to best quantify the impacts of these wastes in LCA. Do these wastes actively leak into the environment? Should studies consider sporadic impacts from large accidental releases of toxic waste? What is the end fate of wastes if the management structure collapses after a century of active management? How questions like these are answered may lead to impacts ranging up to eight orders of magnitude for landfill metal emissions.[92] To date, there is no consensus on whether long-term emissions should be considered in LCA. A report from ecoinvent gives a detailed discussion of the arguments for, and against, including long-term emissions in LCA. Further, the ecoinvent database includes two versions of each impact method ("w/ LT" including long-term emissions and "w/o LT" ignoring the emissions) to help address this issue.[93] Although it is unclear whether, and how, long-term emissions should be quantified, these decisions can greatly influence conclusions about the sustainability of a system (e.g., for waste remediation and accumulation).

## 3.3   Uncertainty, Sensitivity, and Scenario Analysis

Sensitivity analysis quantifies the relationship between input uncertainty and model outputs (e.g., sustainability metrics like NPV and GW impact), which is essential for establishing defensible conclusions from a model. In the LCA and TEA studies we reviewed, primarily, local sensitivity analysis was performed, where parameters were individually varied (typically ± 20%) to quantify how each parameter affected sustainability metrics. However, these local sensitivity analyses did not define realistic ranges of input uncertainty nor consider the possibility of interaction between parameters. Global sensitivity methods can vary multiple parameters simultaneously from well-defined uncertainty distributions (e.g., Monte Carlo),[94–96] but these global methods were rarely performed in the reviewed studies (3 of 16 TEA studies). Instead, most studies performed scenario analyses that determined the range of possible outcomes using best- and worst-case scenarios of parameters (e.g., technology performance and market prices). Future works must go beyond varying individual parameters to define a range of outcomes. Using global methods, we can more rigorously quantify the likelihood of achieving a specific sustainability goal under uncertainty. This probabilistic understanding of system sustainability can be used to more clearly identify research targets for influential parameters, help decision makers understand uncertainties in the model, and identify research direction in large decision spaces.

## 3.4  Disconnection between Economic and Environmental Assessments and the Emerging Consideration of Social Impacts

By understanding the tradeoffs between profitability, environmental, and social impact categories over the product lifecycle, decisions can be made early in technology development to optimize system sustainability. In many REE recovery studies, LCA and TEA are conducted separately, often using different assumptions, system boundaries, and contextual values. This disconnection can lead to inconsistencies and even misleading conclusions. For instance, most of the LCA studies of conventional REE production consider the selective separation and refining steps within the scope of the system (Table 2). However, 7 of the 16 TEA studies did not include these process sections. Therefore, very different REE product values were used to determine profitability in TEA and economic allocation in the LCA studies. Further, many of the systems reviewed led to the generation or remediation of waste. The LCA studies made assumptions about how strictly the waste management would be held to regulations, while TEA studies assigned costs to treat or manage these wastes that do not always align with the LCA assumptions (e.g., assume costs or impacts of waste management are negligible). For example, one study determined that most of the revenue of REE production was from the remediation of the waste they used as a feedstock.[22] These are just two examples of how the lack of integration can lead to an incomplete understanding of the overall sustainability of a technology or process.

Such misalignments highlight the need for integrated sustainability assessment, where model inputs to environmental and economic evaluations are harmonized. Aligning assumptions, data sources, and methodological choices across LCA and TEA ensures a more accurate and holistic understanding of trade-offs and synergies, ultimately supporting better-informed decision-making in research, development, and policy.

## 3.5  Emerging Social Impact Assessments

Social life cycle assessment (sLCA) enables early identification of potential social risks and stakeholder concerns. This is particularly critical in the context of REE recovery, as many REEs are traditionally sourced from regions with poor labor practices, environmental justice issues, and other societal challenges. Assessing social sustainability in emerging REE recovery pathways will help to reduce reliance on exploitative supply chains, promote fair labor practices, and support a more equitable resource distribution. Despite its importance, the least considered pillar of sustainability is social impact. Of the critical minerals, REEs are among the least explored in the social sciences despite many stories of sociological harm to communities.[9]  Near Baotou in China, people have cited livestock and crops dying after the start of REE mining in the region, resulting in a decrease in one town's population from 2000 to 300 people within 10 years. [97–99] The Mountain Pass mine in the U.S. had 60 wastewater spills amounting to 600,000 gallons of toxic wastewater entering the environment. Mismanagement here likely led to the opposition of a new REE mine in Minnesota, where 98% of comments from the public were negative.[97,98] In Malaysia, a past radioactive materials leak led to lawsuits claiming environmental damages that were ultimately refuted by the courts. Consequently, when a new REE refining site was permitted in Kuantan, locals had environmental concerns.[97]

These stories highlight some of the sociological issues pertaining to REE production globally. Broadly, in the literature, socio-ecological concerns revolve around three connected issues: engagement of local communities, environmental justice, and the cost of mitigating impacts.[9,97–100] Firstly, indigenous and local communities near REE resources are directly affected by the impacts of REE production and, therefore, should be engaged in the permitting process. However, in many cases in critical mineral production, companies obtain permission from state governments without communicating with the local community. As each culture is unique, failing to engage these local communities has led to loss of livelihood (e.g., farmers), dispossession of land, civil unrest, and, in extreme cases, war.[99] Greater communication with local communities can ensure that wealth generation extends to improving opportunities for these communities. In addition, those affected tend to be from disadvantaged communities, particularly in the Global South where environmental and human health regulations are weaker.[99,100] Strict regulations are commonly framed as making operations less economically competitive. Therefore, it is critical to quantify these tradeoffs between profit, environmental, and social impact to all stakeholders. The conversation about the cost of cleaning up hazardous waste is primarily confined to after the damage has occurred. However, the multigenerational diseases that occur from exposure to these wastes can shift the cost from businesses to individuals and social systems. Therefore, we should prioritize limiting impacts from before and during operation since this period has a limited time horizon and is geographically contained to the site of production thereby decreasing the total cost to businesses and society.[100] However, robust methodologies that quantify potential impacts prior to production are required.

Of the three pillars of sustainability, social sustainability analysis methods are the newest and least developed. There has been significant progress in developing and supporting social impact assessment resulting in the UNEP-SETAC Life Cycle Initiative (sLCA) Framework that is distributed in two documents.[16,17] The sLCA methodology is similar to environmental LCA in that it follows ISO 14040/14044 standards and has familiar phases for the analysis (goal and scope definition, LCI, LCIA, interpretation). The sLCA methodology has primarily been applied to waste management and bio-based product systems.[101] However, there have been a number of studies relating to critical mineral and REE mining.[102–108] Of these studies, only two relate specifically to REO production, though they both originate from the same group of researchers.[105,106] They found that corruption, bribery, and fair competition in REO and magnet production supply chains pose risks to social structures. Both studies found the Mountain Pass and Mount Weld mines have the least social impact while the Chinese and Malaysian systems have the highest impact. However, these studies are limited in several ways. For example, these studies didn't perform interviews with local communities to supplement their data nor actual price data as inputs with the PSILCA database. In addition, missing data in China and Malaysia processes is treated as "low risk", which could lead to an underestimation of impacts for the Bayan Obo and Mount Weld supply chains. Further, since the PSILCA database only supports the activity variable of "worker hours", the relevance of the results is limited to the "worker" stakeholder group.[105,106]

Ultimately, sLCA methods require further development. Specifically, sLCA methodologies must have better standardization to enable comparisons between studies. Data availability and quality need to improve, especially for local site-specific data.[102] To this end, some background databases exist (PSILCA and SHDB) and can be used in common LCA software like openLCA, SimaPro, and GaBi. However, database scopes have limited data and stakeholder groups available, leading to results that may not accurately quantify all social impacts of a system. Further, there is no direct relationship between sLCA impact categories and the UN sustainable development goals leading to incomplete coverage of discrimination issues.[101] Despite these limitations, sLCA has been helpful in identifying potential hotspots in prospective system designs. Highlighting these hotspots as part of the discussion around novel REO production systems can help increase awareness of common social impact vulnerabilities and encourage smarter designs in the future.[98,104]

# 4 Recommendations and Conclusions

This review aimed to develop a clearer understanding of the sustainability of potential REE production pathways. We synthesized research in the field and identified limitations of these TEA, LCA, and sLCA studies. For TEA, we examined a wide range of different feedstocks and technologies. We identified challenges in identifying products and coproducts, variability in REE values causing uncertainty, and observed that OPEX was most influential on profitability (specifically material costs). For LCA, there were a variety of uncertainties related to parameter (data availability), scenario (functional unit and allocation), and model uncertainty (LCIA characterization factors for water depletion and toxic waste management). Though there were few sLCA studies for REEs, we highlighted some of the key risks from REE mining experienced by local communities who bear the burden of REE acquisitions, often with limited engagement in the process. From this review, we recommend the following to advance this field:

**(1) Standardize methodologies for evaluating REE recovery systems.** We have a few specific recommendations to improve the accuracy and consistency of LCA and TEA studies, as outlined in Section 3. In TEA studies, profitability is largely tied to revenue from REEs, which is highly variable. We recommend reporting profitability by the indicator minimum selling price (MSP), as its calculation does not rely on the price of the main product. Therefore, TEA studies will remain relevant regardless of REE price fluctuations allowing direct comparison with future studies. Additionally, low TRL systems can be expected to become more optimized as the technology develops resulting in improved profitability. Not considering this development could disincentivize emerging technologies. The profitability of an $N^{th}$-of-a-kind plant can be estimated based on the cost of the first-of-a-kind plant and a learning rate based on different methodologies (e.g., from NETL, EIA).[109] Some other LCA/TEA methods (e.g., Lang factors, product value identification, co-product handling, and DCF analysis, etc.) as well as characterization factors applied in REE and critical mineral recovery studies, need to be further studied and standardized by researchers in this field to improve consistency and methodological rigor.

In LCA studies, we recommend all studies report a common mREE or sREE functional unit and the REE composition for enhanced comparability. If a study has a specific end-use for its product, then an additional functional unit and allocation can be used for that specific application (e.g., iREO, Nd for magnets). Further, harmonization of studies for conventional REE systems[58] should be completed with more rigorous uncertainty and sensitivity analysis, as outlined in point 3 below.

**(2) Enable comprehensive assessment and understanding of impacts from radioactivity, water use, and waste handling and treatment.** Although radioactivity, water use, and waste management are regulated by governments and significantly influence the profitability of REE production systems, they are infrequently included within the scope of literature studies. This exclusion reflects the limited inventory and cost data, as well as the absence of established standards or best-practices for quantifying waste management costs and environmental impacts of REE systems. We recommend that future profitability assessments include both capital and operational expenses related to waste management. To quantify environmental impacts using current LCIA methods, we suggest applying ReCiPe 2016 for radioactive substances and AWARE for water depletion. However, further methodological development and

improved completeness are needed to enable more robust impact assessments related to radioactivity, water use, and waste. Improving data availability and advancing LCIA methodologies will be essential for developing a comprehensive understanding of the costs and environmental impacts associated with REE production.

**(3) Account for uncertainties and expand scenario analysis.** Global scenario, uncertainty, and sensitivity analysis methods that simultaneously consider uncertainties from parameters (e.g., reactor temperature or electricity price) and scenarios (e.g., functional unit of iREO vs. sREO) are essential for comprehensive sustainability assessments. Comparing one number to another ignores the complexity of these systems and leads to erroneous conclusions (e.g., GW impact of ethanol fuels). Stochastic methods, like Monte Carlo analysis, have been used in other fields and are equipped to evaluate the entire design space. As opposed to varying parameters by ± 20%, physically reasonable limits should be used to bound parameters and understand indicator sensitivity. Targets for key parameters can be identified based on different profitability and environmental goals to more clearly guide future research efforts.[110] In addition, methodological studies should explore how including model uncertainty (e.g., LCIA methods, long-term impacts) may influence conclusions on the sustainability of these REE systems to avoid pursuing unsustainable paths.[111]

**(4) Integrate life cycle assessment and techno-economic analysis and incorporate social sustainability into this field**. Combining environmental and economic evaluations within a unified framework can provide a more holistic understanding of system performance and trade-offs. Social impacts remain underexplored but are critical in REE supply chains. Including social life cycle assessment (S-LCA) enables early identification of risks and supports fair, responsible sourcing of REEs. Ideally, TEA, LCA, and sLCA would be completed together for greatest value. However, sLCA is still a developing methodology. Therefore, we recommend that TEA and LCA studies discuss the potential social risk factors associated with their systems qualitatively. With good data availability, future sLCA studies can leverage these data to quantify the social impacts and hotspots of emerging systems.

In summary, this review has identified key challenges in the sustainability assessments of REE production systems and highlighted the needs to standardize the approaches for enabling comparisons between systems. Further, as more data and studies become available in this field, we expect that solutions to the above-mentioned challenges will become clearer. These system-level sustainability analyses will be instrumental in ensuring the supply of REEs for consumer and clean energy technologies in the present, while supporting long-term resource availability and sustainability for future generations.

## Conflict of Interest
The authors report no conflict of interest.

## Acknowledgements
We thank the National Science Foundation for funding this research through award number ECO-CBET-2133530.24

# Toward Sustainable Rare Earth Element Production: Key Challenges in Techno-Economic, Life Cycle, and Social Impact Assessment

Adam Smerigan[1] and Rui Shi[1,2]*

[1]Department of Chemical Engineering, The Pennsylvania State University, University Park, Pennsylvania 16802, United States

[2]Institute of Energy and the Environment, the Pennsylvania State University, University Park, Pennsylvania 16802, United States

*Corresponding author: rms6987@psu.edu

## Contents

# 1 Calculation of the Adjustment Factor for Illegal Mining of REEs

Since impacts from illegal REE mining are significant, we suggest that it should be considered when comparing to the environmental impacts of developing systems in other regions. To estimate the additional impacts of illegal REE production in China, we calculated an "adjustment factor". This adjustment factor used literature data on the quantified impacts of illegal REE mining and the capacity of illegal mining to estimate how much additional impact could be attributed to Chinese REE production. For ease of use, we averaged this adjustment factor across all impact categories and production routes. However, a more accurate analysis could be done using this method for individual impact categories and production routes. To use this adjustment factor, any environmental impact from conventional REE mining in Chinese production routes can be multiplied by the adjustment factor to estimate the additional impact from illegal mining in these regions.

Table S1: Ratio of environmental impacts from illegal mining to legal mining for three production routes. Original data from Lee and Wen 2018.[1]

|  | Production Route | | | |
| --- | --- | --- | --- | --- |
| Impact Category | Bayan Obo | Sichuan | Southern Provinces | Average |
| Acidification | 5.88 | 1.34 | 1.89 | |
| Ecotoxicity | 1.57 | 1.60 | 1.54 | |
| Eutrophication | 5.72 | 4.63 | 2.54 | |
| Fine Dust | 2.10 | 1.35 | 1.80 | |
| Human Toxicity | 1.68 | 1.41 | 1.51 | |
| Global Warming | 1.89 | 1.37 | 1.71 | |
| Photochemical Oxidation | 2.23 | 1.18 | 1.47 | |
| Chinese Abiotic Depletion Potential | 2.01 | 1.42 | 1.66 | |
| Average across categories | 2.89 | 1.79 | 1.76 | 2.15 |
| Standard Deviation | 1.81 | 1.15 | 0.35 | |

Table S2: Calculation of the factor used to adjust the impacts of legal REE mining in China considering the additional impacts of illegal mining at these locations.

| Parameter | Value |
| --- | --- |
| Percentage of REOs that are produced illegally (average of scenarios from Lee and Wen - 2018) | 17 |
| Ratio of illegal to legal mining impacts (averaged across production routes) | 2.15 |
| Impact from illegal mining | 0.17 * 2.15 = 0.359 |
| Impacts of legal mining | 1 - 0.17 = 0.83 |
| Adjustment factor for illegal REE mining | 0.359 + 0.83 = 1.19 |

## 2 Compiled LCA Data

Table S3: Summary of LCA studies reviewed. Locations for conventional REE production are Bayan Obo (BO), China Southern Provinces (SP), Mount Weld (MW), and Mountain Pass (MP).

| Author Name | Reference | Date | Feedstock | Paper Focus | Location | Functional Unit |
|---|---|---|---|---|---|---|
| Arshi | 2 | 2018 | Conventional | Analysis | BO, SP | iREO, iREM, REEp |
| Bailey | 3 | 2020 | Conventional | Review | BO, SP, MW | sREO |
| Marx | 4 | 2018 | Conventional | Analysis | BO, MW, MP | REEp |
| Zaimes | 5 | 2015 | Conventional | Analysis | BO | mREO |
| Deng | 6 | 2019 | Conventional | Analysis | SP | mREO |
| Lee | 7 | 2017 | Conventional | Analysis | BO, SP | iREO, sREO |
| Lee | 8 | 2018 | Conventional | Harmonization | BO, SP, SC | sREM |
| Zapp | 9 | 2018 | Conventional | Analysis | BO, SP, NK | iREO, sREO |
| Zapp | 10 | 2022 | Conventional | Analysis | all | iREO |
| Vahidi | 11 | 2016 | Conventional | Analysis | SP | mREO |
| Vahidi | 12 | 2017 | Conventional | Separation | BO, SP | mREE |
| Vahidi | 13 | 2018 | Conventional | Refining | SP | iREF, iREM |
| Schreiber | 14 | 2016 | Conventional | Analysis | BO, NK | iREM |
| Schreiber | 15 | 2021 | Conventional | Review | BO, SP | iREO, sREO |
| Koltun | 16 | 2014 | Conventional | Analysis | BO | iREO |
| Koltun | 17 | 2020 | Conventional | Analysis | MW | iREO, sREO |
| Schulze | 18 | 2017 | Conventional | Analysis | SP | sREO |
| Sprecher | 19 | 2014 | Both | Analysis | BO | REEp, sREO |
| Brown | 20 | 2024 | Secondary | Analysis | US | Soil processed |
| Ippolito | 21 | 2021 | Secondary | Analysis | Italy | Fluorescent powder processed |
| Kulczycka | 22 | 2016 | Secondary | Analysis | Poland | PG treated |
| Alipanah | 23 | 2020 | Secondary | Analysis | US | mREO |
| Rabbani | 24 | 2024 | Secondary | Analysis | US | mREOH, sREO |
| Chowdhury | 25 | 2021 | Secondary | Analysis | US | mREO |
| Jin | 26 | 2016 | Secondary | Analysis | - | REEp |
| Li | 27 | 2019 | Secondary | Analysis | US | Gold |
| Sanchez Moran | 28 | 2024 | Secondary | Analysis | US | mREO |
| Van Nielen | 29 | 2024 | Secondary | Analysis | Europe | REEp |
| Liu | 30 | 2020 | Secondary | Analysis | US | mREO, lighting service |
| Thompson | 31 | 2018 | Secondary | Analysis | US | FCC waste processed |
| Magrini | 32 | 2022 | Secondary | Analysis | Sweden | mREO |

Table S4: Further information about the LCA studies reviewed. If mREEs are separated into sREO products, an "X" is marked in the "separate" column. If sREOs are refined into sREM products, an "X" is marked in the "refine" column.

| Author Name | Reference | Coproduct | Allocation | separation | refining | Hot Spot Analysis | LCI database |
|---|---|---|---|---|---|---|---|
| Arshi | 2 | Iron ore, REOs | E | X | X | X | EI, other |
| Bailey | 3 | Iron ore | E, S | X | | X | EI, Thinkstep |
| Marx | 4 | Iron ore, REOs | E, M | X | X | X | EI 3.3 |
| Zaimes | 5 | Iron ore, REOs | E, M, O | | | | EI 3 |
| Deng | 6 | - | - | | | X | EI 3 |
| Lee | 7 | Iron ore, REOs | E, M | X | X | X | CLCD |
| Lee | 8 | Iron ore | - | X | X | | CLCD |
| Zapp | 9 | Iron ore, REOs | E, S | X | X | X | EI 3.3 |
| Zapp | 10 | Iron ore, REOs | n.a. | X | X | X | n.a. |
| Vahidi | 11 | - | E | | | | EI 3.0 |
| Vahidi | 12 | REOs | E | X | | X | EI 3.0 |
| Vahidi | 13 | REOs | E | X | X | X | EI 3.0 |
| Schreiber | 14 | Iron ore, REOs | E, S | X | X | X | GaBi6, EI 2.2 |
| Schreiber | 15 | - | - | X | | | - |
| Koltun | 16 | Iron ore, REOs | E, M | X | | X | EI |
| Koltun | 17 | REOs | E | X | | | EI |
| Schulze | 18 | - | - | X | | | EI 3.2 |
| Sprecher | 19 | Iron ore, REOs | E | X | X | | EI 2.2 |
| Brown | 20 | - | - | | | | AGRI-BALYSE |
| Ippolito | 21 | mREO | D | | | | Gabi v8.7 |
| kulczycka | 22 | anhydrite, H3PO4 | D | | | | - |
| Alipanah | 23 | REOs | E | | | | - |
| Rabbani | 24 | - | - | X | | X | EI |
| Chowdhury | 25 | Iron oxide | E | | | X | EI 3.7 |
| Jin | 26 | - | - | | | | EI 2.0 |
| Li | 27 | Silver, copper, mREO | E | | | X | EI 3 |
| Sanchez Moran | 28 | Iron salt | E, M | | | X | EI |
| Van Nielen | 29 | - | - | | X | X | EI 3.8, cutoff |
| Liu | 30 | mercury | M, D | | | X | EI 3.5 |
| Thompson | 31 | Eliminated waste disosal | - | | | X | EI 3.0 |
| Magrini | 32 | - | E, M | | | X | EI 3.4 |

## 3 Compiled TEA data

Table S5: General information about the TEA studies reviewed. Some studies have multiple rows because they consider multiple scenarios.

| Source | Unique Information | Feedstock | Capacity | Capacity unit | Capacity | Capacity unit | REO recovery (% mass) |
|---|---|---|---|---|---|---|---|
| [31] | bioleaching | FCC | 18838 | ton FCC catalyst feedstock/year | 136.9523 | mt REO/year | 28-56 |
| [28] | acid-free dissolution | HDD | 342.42 | tonnes of HDD shreds/year | 2.53 | tonnes didymium oxide/year | 75 |
| [33] | | NiMH | 2000 | tonnes/year of used battery powder | 244.6 | t REO concentrate/year | 85.13 |
| [34] | | Cell phone | 2000 | tonnes/year of cell phone waste | 31.71 | t/yr REO concentrate | 78 |
| [32] | bioleaching | NdFeB | 0.33-1.2 | tonnes/year of NIB magnets | 0.095-0.35 | kg/year REO | 83 |
| [25] | acid-free dissolution | NdFeB | 100 | tonnes/year NdFeB magnet sward | 32 | tonnes REO/year | 97 |
| [35] | metal distillation | NdFeB | 744.6 | kg of mischmetal (Nd70) | 215.934 | tonne/year Misch and Dy metal | 60-100 |
| [36] | supercritical fluid extraction | NdFeB | 1228-2266 | tonne/year preprocessed magnet | 302-689 | tonne/year iREO | 97 |
| [36] | supercritical fluid extraction | lamp phosphors | 2606-10987 | tonne/year preprocessed lamp phosphors | 421-2410 | tonne/year iREO | 66 |
| [37] | magnetic nanoparticles | geothermal brine | 1344.375 | tonne/year brine | 2.151 | tonne/year Eu | 90 |
| [38] | detailed SX process for iREO recovery | monazite | 16000 | tonne/year monazite concentrate | 7291 | tonne iREO/year | 88.5 |
| [38] | | monazite | 16000 | tonne/year monazite concentrate | 7114 | tonne iREO/year | 87.2 |
| [38] | | monazite | 16000 | tonne/year monazite concentrate | 7412 | tonne iREO/year | 90 |
| [23] | biosorption using Si-sol gel | coal fly ash | 200000 | tonnes/year CFA | 47.1 | tonnes mREO/year | 70 |
| [23] | | lignite | 200000 | tonnes/year lignite | 83 | tonnes mREO/year | 75 |
| [39] | supercritical fluid extraction | coal fly ash | 3650-4867 | tons/year coal ash | 1.1-2.69 | tons REE/year | 45-60 |
| [40] | | AMD | 4397394 | mt/year acidic mine wastewater | 27 | mt/year mREO | 80 |
| [41] | | AMDp (Sterrett) | 748.7493 | tonne AMDp/year | 1 | tonne/year REE | 80 |
| [41] | | AMDp (Woodlands) | 1133.571 | tonne AMDp/year | 1 | tonne/year REE | 80 |
| [42] | modelled SX, network sourcing strategy | AMDp | 53000 | tonnes AMD (preconcntrate)/year | 444 | tonnes REE/year | 93 |
| [42] | | AMDp | 53000 | tonnes AMD (preconcntrate)/year | 290 | tonnes REE/year | 61 |
| [42] | | AMDp | 53000 | tonnes AMD (preconcntrate)/year | 290 | tonnes REE/year | 61 |
| [42] | | AMDp | 53000 | tonnes AMD (preconcntrate)/year | 290 | tonnes REE/year | 61 |
| [42] | | AMDp | 53000 | tonnes AMD (preconcntrate)/year | 212 | tonnes REE/year | 44 |
| [43] | | PA sludge | 453000 | tonnes PA-sludge/year | 138 | ton REO/year | 48 |
| [43] | | PA sludge | 453000 | tonnes PA-sludge/year | 49 | ton REO/year | 17 |

Table S6: System costs (CAPEX and OPEX), revenues, and profitability as payback period (PBP), return on investment (ROI), net present value (NPV), and internal rate of return (IRR). Information not provided in a paper is listed as a "-" in the table.

| Source | Revenue (M$/year) | CAPEX (M$) | OPEX (M$/year) | PBP | ROI (%) | NPV (M$) | IRR (%) |
|---|---|---|---|---|---|---|---|
| 31 | 3.88 | 1.27 | 1.76 | - | - | 5.78 | 44 |
| 28 | 0.38 | 0.268 | 0.22 | - | - | - | - |
| 33 | 33.7 | 26.92 | 13.88 | 1.58 | 63 | 95.9 | 46.1 |
| 34 | 77.2 | 53.2 | 18.23 | 1.09 | 91.61 | 296 | 60.9 |
| 32 | 0.031 | - | 0.11 | - | - | - | - |
| 25 | 1.24-2.69 | 0.48-0.91 | 0.61 | - | - | - | - |
| 35 | 23.8 | 1.62 | 5.63 | - | - | 299 | 143 |
| 36 | 17-65 | 13.8 | 13-22 | 0.5 | - | 17.5 | 16 |
| 36 | 7.0-58 | 14.8-15.8 | 8.0-23 | 0.5 | - | 128.3 | 49 |
| 37 | 2.2 | 6.8 | 1 | - | - | - | 18.1 |
| 38 | 131 | 90 | 88 | 4.5 | - | 135 | - |
| 38 | 104 | 90 | 99 | - | - | -15 | - |
| 38 | 113.8 | 88 | 101 | - | - | -62 | - |
| 23 | 0.75 | 1.4 | 76.69 | - | - | - | - |
| 23 | 28.55 | 1.2 | 24.40 | - | - | 28 | - |
| 39 | 0.77-2.37 | 0.4-0.5 | 1.3-3.5 | - | - | - | - |
| 40 | 0.783 | 3.835 | 5.33 | - | - | - | - |
| 41 | 0.3 | 22 | 3.4 | - | - | - | - |
| 41 | 0.6 | 16.6 | 5.7 | - | - | - | - |
| 42 | 49.42-70.46 | 186 | 21.4 | - | - | (-80)-270 | 0-32 |
| 42 | 35.26-56.07 | 131 | 17.5 | - | - | (-105)-110 | (-15)-24 |
| 42 | 20.25-33.39 | 149 | 18.5 | - | - | (-100)-160 | (-12)-27 |
| 42 | 20.25-33.39 | 155 | 20.5 | - | - | (-60)-275 | 2.-37 |
| 42 | 49.42-79.83 | 142 | 20.5 | - | - | (-80)-190 | (-60)-31 |
| 43 | 122 | 7.08 | 9.8 | - | - | 442 | |
| 43 | 112 | 7.55 | 27.1 | - | - | 179 | |

Table S7: The discounted cash flow analysis parameters for the reviewed TEA studies. Information not provided in a paper is listed as a "-" in the table. Depreciation methods are straight line (SL), double declining balance (DDB), modified accelerated cost recovery system (MACRS), and declining balance with a depreciation rate of 150% (150% DB).

| Source | Discount rate (%) | Plant life (yr) | Operating time (hrs/yr) | Loan (yr) | Loan interest rate (%) | Depreciation period (yr) | Depreciation method | Income tax rate (%) | Startup time (months) | Construction (yr) | Cost basis |
|---|---|---|---|---|---|---|---|---|---|---|---|
| 31 | 8 | 30 | 8000 | 10 | 8 | 10 | DDB + SL | 35 | 6 | no | - |
| 28 | 10 | 20 | 7884 | 10 | 7.5 | 7 | DDB | 39 | no | no | 2022 |
| 33 | 7 | 15 | 2000 | 10 | 9 | 10 | SL | 25 | 4 | 2.5 | - |
| 34 | 9 | 15 | 2000 | 10 | 9 | 10 | SL | 25 | 4 | 2.5 | - |
| 32 | - | - | - | - | - | 10 | SL | 30 | - | - | - |
| 25 | - | 20 | 2680 | - | - | - | - | - | - | - | - |
| 35 | 3.03 | 30 | 7446 | 15 | 4.5 | 20 | DDB | 35 | 0 | 3 | 2012 |
| 36 | 7.5 | 40 | 7446 | - | - | - | - | - | 0 | 2 | 2023 |
|  | 7.5 | 40 | 7446 | - | - | - | - | - | 0 | 2 | 2023 |
| 37 | 10 | 30 | 8322 | 10 | 4 | 20 | 150% DB | 38 | 0 | 2 | 2018 |
| 38 | 10 | 20 | 8000 | no | no | 20 | SL | 30 | reduced productivity in first two years (80%,90%) | 0 | 2020 |
| 23 | 8 | 20 | 8000 | 10 | 8 | 7, 15, or 19 | MACRS | 27 | 6 | 0 | 2020 |
| 39 | - | 15 | 7300 | - | - | - | - | - | - | - | - |
| 40 | - | 10 | 8760 | - | - | - | - | - | - | - | - |
| 41 | - | 25 | 8000 | - | - | - | - | - | - | - | - |
| 42 | 10 | 20 | - | 10 | 6 | 20 | 150% DB | 26 | 0 | 3 | 2020-2021 |
| 43 | 5 | 10 | - | 0 | 0 | 10 | SL | 35 | 0 | 2 | 2023 |

# 4 Functional Unit Conversion Method

To convert between functional units of mREE, sREO, and sREM, we developed a method that uses literature data from LCAs of conventional REE production. We compiled data from many references to identify the amount of impact from individual process sections (Figure 6e and Figure 2).[2,3,7,9,10,13,17] The data for the REE recovery after further processing of mREEs (Figure 6d) was from one publication[7], as the LCI of other publications were less transparent. The REE content of REO compounds is provided in Table S8 and can be used to determine fractional REE content values dictated by REE composition. These sets of data provide the basis for the proposed conversion method.

Table S8: The REE content of various REO compounds.

| REO | REE Atomic Weight (g/mol) | Mass REO | Mass O | REE Content of REO (wt %) | Deviation from Average REE Content (%) |
|---|---|---|---|---|---|
| $Sc_2O_3$ | 45 | 138 | 90 | 0.65 | 23 |
| $Y_2O_3$ | 89 | 226 | 178 | 0.79 | 7 |
| $La_2O_3$ | 139 | 326 | 278 | 0.85 | 1 |
| $Ce_2O_3$ | 140 | 328 | 280 | 0.85 | 1 |
| $CeO_2$ | 140 | 172 | 140 | 0.81 | 4 |
| $Pr_2O_3$ | 141 | 330 | 282 | 0.85 | 1 |
| $Pr_6O_{11}$ | 141 | 1021 | 845 | 0.83 | 2 |
| $Nd_2O_3$ | 144 | 336 | 288 | 0.86 | 1 |
| $Pm_2O_3$ | 145 | 338 | 290 | 0.86 | 1 |
| $Sm_2O_3$ | 150 | 349 | 301 | 0.86 | 2 |
| $Eu_2O_3$ | 152 | 352 | 304 | 0.86 | 2 |
| $Gd_2O_3$ | 157 | 363 | 315 | 0.87 | 3 |
| $Tb_2O_3$ | 159 | 366 | 318 | 0.87 | 3 |
| $Tb_4O_7$ | 159 | 748 | 636 | 0.85 | 0 |
| $Dy_2O_3$ | 163 | 373 | 325 | 0.87 | 3 |
| $Ho_2O_3$ | 165 | 378 | 330 | 0.87 | 3 |
| $Er_2O_3$ | 167 | 383 | 335 | 0.87 | 3 |
| $Tm_2O_3$ | 169 | 386 | 338 | 0.88 | 3 |
| $Yb_2O_3$ | 173 | 394 | 346 | 0.88 | 4 |
| $Lu_2O_3$ | 175 | 398 | 350 | 0.88 | 4 |
| Average | | | | 0.85 | |

Further, we want to highlight some limitations of this method. This method assumes separation and refining sections of novel systems will perform similarly to conventional systems with different feedstock mREO compositions. Therefore, systems with highly dissimilar compositions of mREE may be less accurate. Further, if a novel system has significantly lower impact in mining, beneficiation, leaching, and concentration sections (the blue bar), the contribution of separation and refining would be larger than conventional routes, especially BO and SP routes. The MW route would be the best choice for systems with low impacts from mREE production. For these three routes, the contributions for are provided in **Error! Reference source not found.**. Another source of inaccuracy is the REE content variable. For the REE content variable, we considered the common oxides formed by REEs and averaged the fractional REE content. This number is within 5% for all REOs except $Y_2O_3$ (7% deviation) and $Sc_2O_3$ (23% deviation). Hence, if a product consists mainly of Sc, a fractional REE content closer to $Sc_2O_3$ (65.2%) may be appropriate.

The most accurate conversion between functional units would to directly add the impact of the separation or refining process sections (using whichever technology) to the mREE impact value (for whichever impact category). However, studies use a wide variety of life cycle impact assessment methods (LCIA), with different units and ways to quantify impact, that make it impossible to add the impact from further processing to the impact of mREE production. In the future, if LCAs of conventional REE production are harmonized, it would be possible to report impacts (with uncertainty) across multiple LCIA methods to enable this alternative method. For now, the proposed method above is suitable to enable rapid comparison of environmental impacts between systems to enhance decision-making.

## 5   Equations for calculating profitability

The following equations were to calculate the profitability, as return on investment (%) and cash flow (M$), using data from the reviewed studies. The calculated profitability was then used for comparisons in Figure 5 of the main text.

Cash Flow = Annualized Revenue – Annualized CAPEX – OPEX

Return on Investment (ROI) = $\frac{\text{(Annualized revenue - Annualized OPEX)}}{\text{CAPEX}} \times 100$